
\input PHYZZX
\title{\bf
Instanton and QCD-monopole Trajectory
\nextline
in the Abelian Dominating System
}

\author{
H.~Suganuma$^{\rm a}$, K.~Itakura$^{\rm b}$,
and H.~Toki$^{\rm a}$
}

\address{
{\rm a)} Research Center for Nuclear Physics (RCNP),
Osaka University,
\nextline
Mihogaoka 10-1, Ibaraki 567, Japan
}

\address{
{\rm b)} College of Arts and Sciences,
Tokyo University,
\nextline
Komaba 3-8-1, Meguro, Tokyo 153, Japan
}

\abstract{
Correlation between instantons and QCD-monopoles is studied
in the abelian-gauge-fixed QCD.
{}From a simple topological consideration,
instantons are expected to appear only around the QCD-monopole
trajectory in the abelian-dominating system.
The QCD-monopole in the multi-instanton solution is studied
in the Polyakov-like gauge, where $A_4(x)$ is diagonalized.
The world line of the QCD-monopole is found to be penetrate
the center of each instanton.
For the single-instanton solution,
the QCD-monopole trajectory becomes a simple straight line.
On the other hand, in the multi-instanton system,
the QCD-monopole trajectory often has complicated topology
including a loop or a folded structure, and is unstable against
a small fluctuation of the location and the size of instantons.
We also study the thermal instanton system in the Polyakov-like gauge.
At the high-temperature limit, the monopole trajectory becomes
straight lines in the temporal direction.
The topology of the QCD-monopole trajectory is drastically changed
at a high temperature.
}


\endpage

\chapter{Abelian Gauge Fixing and QCD-Monopole}

Color confinement is one of the central issues
in the nonperturbative QCD
\REF\kerson{
K.~Huang, {\it Quarks, Leptons and Gauge Fields},
(World Scientific, Singapore, 1982) 1.
}
\REF\rothe{
H.~J.~Rothe, {\it Lattice Gauge Theories}, (World Scientific,
1992) 1.
}
[\kerson,\rothe],
and is characterized by the formation of the color-electric flux tube
[\rothe] with the string tension about 1GeV/fm.
To understand the confinement mechanism,
much attention has been paid for
the analogy between the superconductor and the QCD vacuum
\REF\nambu{
Y.~Nambu, Phys.~Rev.~{\bf D10} (1974) 4262.
}
\REF\thooftA{
G.~'t~Hooft, {\it High Energy Physics},
(Editorice Compositori, Bologna, 1975).
}
\REF\mandelstam{
S.~Mandelstam,  Phys.~Rep.~{\bf C23} (1976) 245.
}
[\nambu-\mandelstam]
using the duality of the gauge theory.
Similar to the superconductivity,
the color-electric flux seems to be
excluded in the QCD vacuum, which leads the formation of the
squeezed color-flux tube between color sources.
In this analogy, color confinement is brought by the dual
Meissner effect originated from color-magnetic monopole condensation,
which corresponds to Cooper-pair condensation in the superconductivity.
As for the appearance of color-magnetic monopoles in QCD,
't~Hooft
\REF\thooftB{
G.~'t~Hooft, {\it Nucl.~Phys.}~{\bf B190} (1981) 455.
}
[\thooftB] proposed an interesting idea
of the abelian gauge fixing, which is defined by
the diagonalization of a gauge-dependent variable $X(x)$.
In this gauge, the nonabelian gauge theory like QCD
is reduced into an abelian gauge theory with QCD-monopoles,
which appear from the hedgehog-like configuration
\REF\sugaA{
H.~Suganuma, S.~Sasaki and H.~Toki,
{\it Nucl.~Phys.}~{\bf B435} (1995) 207.
}
\REF\sugaB{
H.~Suganuma, H.~Ichie, S.~Sasaki and H.~Toki,
Prog.~Theor.~Phys.(Suppl.) (1995) in press.
}
\REF\sugaC{
H.~Suganuma, H.~Ichie, S.~Sasaki and H.~Toki,
{\it Color Confinement and Hadrons},
(World Scientific, 1995) in press.
}
\REF\sugaX{
H.~Suganuma, K.~Itakura, H.~Toki and O.~Miyamura,
Proc. of Int.~Workshop on
{\it Non-perturbative Approach to QCD}, July, 1995,
ECT$^*$, Trento, in press.
}
\REF\sugamiya{
H.~Suganuma, A.~Tanaka, S.~Sasaki and O.~Miyamura,
Proc.~of~Int.~Symposium on {\it Lattice Field Theory},
Melbourne, 1995, Nucl.~Phys.~{\bf B} in press.
}
[\sugaA-\sugamiya]
corresponding to the nontrivial homotopy class
on the nonabelian manifold,
$\pi _2({\rm SU}(N_c)/{\rm U}(1)^{N_c-1})=Z_\infty ^{N_c-1}$.

To begin with, the abelian gauge fixing
is studied with attention to the ordering condition
[\sugaX,\sugamiya],
which is closely related to the magnetic charge of QCD-monopoles.
In general, the abelian gauge fixing consists of two sequential
procedures.

\item{1.} The diagonalization of a gauge-dependent variable
$X(x)$ by a suitable gauge transformation : $X(x)\rightarrow X_d(x)$
[\sugaA].
Since there also remains a discrete symmetry corresponding to
the permutation of the diagonal elements of $X_d(x)$,
the gauge group SU$(N_c)_{\rm local}$ is reduced to
U(1)$^{N_c-1}_{\rm local} \times P^{N_c}_{\rm global}$
by the diagonalization of $X(x)$.

\noindent
$P^{N_c}$-symmetry here becomes a global symmetry
in the description with the continuous field theory, where
local discrete changes are forbidden.
On the other hand, such a $P^{N_c}$-symmetry often appears as
a local symmetry in the lattice gauge theory
in case of the abelian gauge fixing without the ordering condition,
which may lead to a problematic ultra-violet behavior of the field
variable.

\item{2.} The ordering on the diagonal elements of $X_d(x)$
by imposing the additional condition, for instance,
$$
X_d^1(x) \ge X_d^2(x) \ge ... \ge X_d^{N_c}(x).
\eqn\ONE
$$
The residual gauge group U(1)$^{N_c-1}_{\rm local}
\times P^{N_c}_{\rm global}$
is reduced to U(1)$^{N_c-1}_{\rm local}$
by the ordering condition on $X_d(x)$ [\sugaX,\sugamiya].

The magnetic charge of the QCD-monopole is closely related to
the ordering condition in the diagonalization in the abelian
gauge fixing [\sugaX,\sugamiya].
For instance,
in the SU(2) case,
the hedgehog configuration as $X(x) = ({\bf x} \cdot \tau )$ and
the anti-hedgehog one as $X(x) = -({\bf x} \cdot\tau )$
provide a QCD-monopole with an opposite magnetic charge,
``anti-QCD-monopole'',
because they are connected by the additional gauge transformation by
$$
\Omega =\exp\{i\pi ({\tau ^1 \over 2} \cos\alpha +{\tau ^2 \over 2} \sin\alpha
)\}
\in P^2_{\rm global}
\eqn\TWO
$$
with an arbitrary constant $\alpha $.
Here, $\Omega $ physically means the rotation of angle $\pi $
in the internal SU(2) space, and it
interchanges the diagonal elements of $X_d(x)$, which leads
a minus sign in the U(1)$_3$ gauge field, $A_\mu ^3(x)$.
Thus, the magnetic charge of the QCD-monopole is settled
by imposing the ordering condition on $X_d(x)$.

$P^{N_c}_{\rm global}$-symmetry is also important for the argument of
gauge dependence.
If a variable holds the residual gauge symmetry in the abelian gauge,
it is proved to be SU($N_c$) gauge invariant
\REF\hioki{
S.~Hioki, S.~Kitahara, S.~Kiura, Y.~Matsubara,
O.~Miyamura, S.~Ohno and
\nextline
T.~Suzuki, Phys.~Lett.~{\bf B272} (1991) 326.
}
[\hioki].
However, one should carefully examine the residual gauge symmetry,
which often includes not only U(1)$^{N_c-1}_{\rm local}$
but also $P^{N_c}_{\rm global}$.
For instance, the dual Ginzburg-Landau theory [\sugaA] is,
strictly speaking, an effective theory
holding U(1)$^{N_c-1}_{\rm local} \times P^{N_c}_{\rm global}$
symmetry.
Hence, gauge dependence of a physical variable
should be carefully checked in terms of the residual gauge symmetry
U(1)$^{N_c-1}_{\rm local} \times P^{N_c}_{\rm global}$
instead of U(1)$^{N_c-1}_{\rm local}$ [\hioki].
As a result, the dual gauge field $\vec B_\mu $ is not
SU($N_c$)-invariant,
because $\vec B_\mu $ is U(1)$^{N_c-1}$-invariant but
is changed under the global $P^{N_c}$ transformation.

We briefly compare the dual Higgs mechanism in the nonperturbative
QCD vacuum with the ordinary Higgs mechanism.
Like the Cooper pair in the superconductivity
or the Higgs field in the standard theory [\kerson],
the charged-matter field to be condensed
is the essential degrees of freedom for the Higgs mechanism.
On the other hand, there is only the gauge field in the
pure gauge QCD, and hence it seems difficult to find any
similarity with the Higgs mechanism.
In the abelian gauge, however, only the diagonal gluon behaves
as the gauge field, and the off-diagonal gluon behaves
as the charged-matter field, which leads QCD-monopoles
as the relevant degrees of freedom for color confinement.
Condensation of QCD-monopoles leads to mass generation
of the dual gauge field through the dual Higgs mechanism
\REF\suzuki{
T.~Suzuki, {\it Prog.~Theor.~Phys.}~{\bf 80} (1988) 929 ;
{\bf 81} (1989) 752. \nextline
S.~Maedan and T.~Suzuki, {\it Prog.~Theor.~Phys.}~{\bf 81} (1989) 229.
}
[\sugaA,\suzuki],
and therefore the QCD vacuum would be regarded
as the dual superconductor by the abelian gauge fixing.
In this framework, the nonperturbative QCD is mainly
described by the abelian gauge theory with QCD-monopoles,
which is called as the abelian dominance
\REF\iwasaki{
Z.~F.~Ezawa and A.~Iwazaki, Phys.~Rev.~{\bf D25} (1982) 2681.
\nextline
Z.~F.~Ezawa and A.~Iwazaki, Phys.~Rev.~{\bf D26} (1982) 631.
}
[\thooftB,\iwasaki].

Many recent studies
\REF\kronfeld{
A.S.~Kronfeld, G.~Schierholz and U.-J.~Wiese,
Nucl.~Phys.~{\bf B293} (1987) 461.
}
\REF\suzuki{
T.~Suzuki and I.~Yotsuyanagi,
Phys.~Rev.~{\bf D42} (1990) 4257.
}
\REF\miyamura{
O.~Miyamura, Nucl.~Phys.~{\bf B}(Proc.~Suppl.){\bf 42} (1995) 538.
}
\REF\shiba{
H.~Shiba and T.~Suzuki, Phys.~Lett.~{\bf B333} (1994) 461.
}
\REF\kitahara{
S.~Kitahara, Y.~Matsubara and T.~Suzuki, Prog.~Theor.~Phys. {\bf 93}
(1995) 1.
}
\REF\ejiri{
S.~Ejiri, S.~Kitahara, Y.~Matsubara and T.~Suzuki,
Phys.~Lett.~{\bf B343} (1995) 304.
}
\REF\origuchi{
O.~Miyamura and S.~Origuchi,
{\it Color Confinement and Hadrons},
(World Scientific, 1995) in press.
}
\REF\woloshyn{
R.~M.~Woloshyn, Phys.~Rev.~{\bf D51} (1995) 6411.
}
\REF\giacomo{
A.~Di Giacomo,
Proc.~of~Int.~Symposium on {\it Lattice Field Theory},
Melbourne, 1995, Nucl.~Phys.~{\bf B} in press.
}
[\sugamiya,\hioki, \kronfeld-\giacomo]
based on the lattice gauge theory
have supported the realization of QCD-monopole condensation
and the abelian dominance on the color confinement
and other nonperturbative quantities of QCD
in the maximally abelian gauge and/or in the Polyakov gauge.
The crucial role of QCD-monopole condensation
to the chiral-symmetry breaking is also supported
by recent lattice studies [\miyamura,\origuchi,\woloshyn]
and the model analyses
\REF\sugaD{
H.~Suganuma, S.~Sasaki and H.~Toki,
{\it Quark Confinement and Hadron Spectrum},
Como, Italy, (World Scientific, 1995) p.238.
\nextline
S.~Sasaki, H.~Suganuma and H.~Toki, {\it ibid} p.241.
}
\REF\sasaki{
S.~Sasaki, H.~Suganuma and H.~Toki, Prog.~Theor.~Phys. {\bf 94} (1995) 373.
}

[\sugaA-\sugaC,\sugaD,\sasaki].

In this paper, we study the relation between instantons
\REF\rajaraman{
R.~Rajaraman, {\it Solitons and Instantons},
(North-Holland, Amsterdam, 1982) 1.
} [\rajaraman] and QCD-monopoles in the abelian gauge.
The instanton is also an important topological object [\rajaraman]
in the nonperturbative QCD appearing in the Euclidean 4-space
corresponding to the nontrivial homotopy class on the nonabelian
manifold, $\pi _3({\rm SU}(N_c))$= $Z_\infty $.
If the system is completely described only by the abelian field,
the instanton would lose the topological basis
for its existence, and therefore it seems unable to
survive in the abelian manifold.
However, even in the abelian gauge, nonabelian components remain
relatively large around the QCD-monopoles, which are nothing
but the topological defects, so that instantons
are expected to survive only around the world lines of
the QCD-monopole in the abelian-dominating system.
The close relation between instantons and QCD-monopoles are thus
suggested from the topological consideration.

\chapter{
QCD-monopoles in the Multi-Instanton System
}

We demonstrate a close relation between instantons
and QCD-monopoles in the Euclidean SU(2) gauge theory in continuum
[\sugaA-\sugaD].
Since there is an ambiguity on the
gauge-dependent variable $X(x)$ to be diagonalized
in the abelian gauge fixing [\thooftB,\sugaA],
it would be a wise way to choose a suitable $X(x)$
so that the instanton configuration can be simply described.
Here, we adopt the Polyakov-like gauge [\sugaX,\sugamiya],
where $A_4(x)$ is diagonalized.
The Polyakov-like gauge has a large similarity to the Polyakov gauge,
because the Polyakov loop $P(x)$ is also diagonal in this gauge.
%

Using the 't~Hooft symbol $\bar \eta ^{a\mu \nu }$,
the multi-instanton solution is written as
[\rajaraman]
$$
\eqalign{
A^\mu (x) &=i\bar \eta ^{a\mu \nu }{\tau ^a \over 2} \partial^\nu \ln \phi (x)
        =-i{\bar \eta ^{a\mu \nu }\tau ^a \over \phi (x)}
\sum_k {a_k^2 (x-x_k)^\nu  \over |x-x_k|^4},
\cr
\phi (x)  &\equiv 1+\sum_k {a_k^2 \over |x-x_k|^2},
}
\eqn\THREE
$$
where $x_k^\mu  \equiv ({\bf x}_k,t_k)$ and $a_k$ denote the center
coordinate and the size of $k$-th instanton, respectively.
In this case, one finds
$$
A_4(x) =i{\tau ^a \over \phi (x)}
\sum_k {a_k^2 ({\bf x}-{\bf x}_k)^a \over |x-x_k|^4},
\eqn\THREEhalf
$$
and therefore, near the center of $k$-th instanton,
$A_4(x)$ takes a hedgehog configuration around ${\bf x}_k$,
$$
A_4(x) \simeq i {\tau ^a ({\bf x}-{\bf x}_k)^a \over |x-x_k|^2}
\eqn\FOUR
$$
like a single-instanton solution.
In the Polyakov-like gauge,
$A_4(x)$ is diagonalized by a singular gauge transformation,
which provides a QCD-monopole on the center of the hedgehog,
${\bf x} = {\bf x}_k$.
Thus, the center of each instanton is inevitably penetrated
by a QCD-monopole trajectory along the temporal
direction in the Polyakov-like gauge [\sugaC,\sugaX,\sugamiya].
In other words, instantons exist only along the
QCD-monopole trajectories.

For the single-instanton system,
$A_4(x)$ takes a hedgehog configuration around ${\bf x}_1$,
$$
A_4(x)=i a_1^2 {\tau ^a ({\bf x}-{\bf x}_1)^a \over
(x-x_1)^2 \cdot \{ (x-x_1)^2+a_1^2 \} }.
\eqn\FIVE
$$
The diagonalization of $A_4(x)$ is carried out using
a time-independent singular gauge transformation
with the gauge function
$$
\Omega ({\bf x})=e^{i\tau _3\phi }\cos {\theta  \over 2}
+{i(\tau _1\cos\alpha +\tau _2\sin\alpha )} \sin {\theta  \over 2}
=
\pmatrix {
e^{i\phi }\cos{\theta  \over 2}     &    ie^{i\alpha }\sin{\theta  \over 2}
\cr
ie^{-i\alpha }\sin{\theta  \over 2}   &    e^{-i\phi }\cos{\theta  \over 2}
}
\eqn\SIX
$$
with $\theta $ and $\phi $ being the polar and azimuthal angles,
$$
{\bf x}-{\bf x}_1=(\sin\theta \cos\phi ,\sin\theta \sin\phi ,\cos\theta ).
\eqn\SIXhalf
$$
Here, $\alpha $ is an arbitrary constant angle corresponding to the
residual U(1)$_3$ symmetry.
Since $\Omega ({\bf x})$ is time-independent, $A_4(x)$ is simply
transformed as
$$
A_4(x)\rightarrow \Omega ({\bf x})A_4(x)\Omega ^{-1}({\bf x}).
\eqn\SEVEN
$$

After the singular gauge transformation by $\Omega ({\bf x})$,
the abelian gauge field $A_\mu ^3(x)$ has a singular part
stemming from
$$
A_\mu ^{sing}(x)
={1 \over e}\Omega ({\bf x}) \partial_\mu  \Omega ^{-1}({\bf x}),
\eqn\EIGHT
$$
which leads to the QCD-monopole with the magnetic
charge $g=4\pi /e$ [\sugaA].
The QCD-monopole appears at the center of the hedgehog,
${\bf x}={\bf x}_1$, which satisfies $A_4(x)=0$ in Eq.{\FIVE}.
Hence, the QCD-monopole trajectory $x^\mu \equiv({\bf x},t)$
becomes a simple straight line penetrating the center of the instanton
as shown in Fig.1 (a),
$$
{\bf x}={\bf x}_1 \hbox{\quad} (-\infty <t<\infty ),
\eqn\NINE
$$
at the classical level in the Polyakov-like gauge
[\sugaA,\sugaC,\sugaX,\sugamiya].
Similar relation for the QCD-monopole in a single instanton
is found also in the maximally abelian gauge
\REF\chernodub{
M.~N.~Chernodub and F.~V.~Gubarev, JETP~Lett. {\bf 62} (1995) 100.
\nextline
A.~Hart, M.~Teper, preprint OUTP-95-44-P (1995), hep-lat/9511016.
}
[\chernodub].

It should be noted that
the singularity of $A_\mu (x)$ at the center of the instanton
can be removed easily by a gauge transformation
to the non-singular gauge [\rajaraman],
where the singular-free gauge field,
$$
A_\mu (x)=i {\tau ^a ({\bf x}-{\bf x}_1)^a \over (x^2-x_1^2)+a_1^2},
\eqn\NINEhalf
$$
provides the same QCD-monopole trajectory as mentioned above.
It is also worth mentioning that
the QCD-monopole trajectory
is not changed by the residual U(1)$_3$-gauge transformation,
so that QCD-monopoles in the Polyakov-like gauge
are identical to those, {\it e.g.}, in the temporal gauge: $A_4(x)=0$.

For the single anti-instanton system,
one has only to replace $A_4(x)\rightarrow -A_4(x)$ corresponding to
$\bar \eta ^{a\mu \nu } \rightarrow  \eta ^{a\mu \nu }$ in the above argument
[\rajaraman].
Since this replacement interchanges the hedgehog and the
anti-hedgehog on $A_4(x)$,
it leads to the change of the QCD-monopole charge
as mentioned in Section 2.1.
Then, the anti-QCD-monopole with the opposite magnetic charge, $-g$,
appears and passes through the center of the anti-instanton
as shown in Fig.1 (b).
In Figs.1 (a) and (b), relative difference on the
QCD-monopole charge is expressed by the direction of the arrow.

For the two-instanton system, two instanton centers
can be put on the $zt$-plane by a suitable spatial rotation
in ${\bf R}^3$ without loss of generality,
so that one can set $x_1=y_1=x_2=y_2=0$.
Owing to the axial-symmetry around the $z$-axis of the system,
the QCD-monopole trajectory only appears on the $zt$-plane, and hence
one has only to examine $A_4(x)$ on the $zt$-plane by setting $x=y=0$.
In this case, $A_4(x)$ is already diagonalized on the $zt$-plane:
$$
A_4(z,t;x=y=0)
= i {\tau^3  \over \phi (z,t)}
\sum_{k=1}^2 a_k^2 {(z-z_k) \over \{ (z-z_k)^2+(t-t_k)^2 \}^2}
\equiv A_4^3(z,t)\tau ^3.
\eqn\TEN
$$
Therefore, the QCD-monopole trajectory $x^\mu =(x,y,z,t)$
is simply given by $x=y=0$ and $A_4^3(z,t)=0$ or
$$
\sum_{k=1}^2 a_k^2 {(z-z_k) \over \{ (z-z_k)^2+(t-t_k)^2 \}^2}=0.
\eqn\ELEVEN
$$
Here, $A_4(x)$ takes a hedgehog or an anti-hedgehog
configuration near the QCD-monopole at each $t$ [\sugaX,\sugamiya].
The shape of the QCD-monopole trajectory depends only on the relative
vector between the instanton centers and
the ratio $a2/a1$ between the instanton sizes, which means
the irrelevance of the absolute value of the instanton size.
The typical scale of the system is mainly characterized by
the relative distance between instantons,
because the classical Yang-Mills theory has no scale-parameter.

We show in Figs. 2 (a),(b) and (c) the typical examples
of the QCD-monopole trajectory in the two-instanton system.
The QCD-monopole trajectories are found to be rather
complicated even at the classical level.
Fig.2 (a) shows the simplest case for two instantons
with the same size, $a_1=a_2$, locating at the same Euclidean time,
$(z_1,t_1)=-(z_2,t_2)=(z_0,0)$.
In this case, the QCD-monopole trajectory
$(z,t)$ is analytically solved [\sugaC,\sugaX] as
$$
z=0
\hbox{\quad or \quad}
t^2=(z_0^2-z^2)+2|z_0|\sqrt{(z_0^2-z^2)},
\eqn\TWELVE
$$
and there appear two junctions and a loop
in the QCD-monopole trajectory [\sugaC,\sugaX].
Here, the QCD-monopole charge calculated
is expressed by the direction of the arrow.
There also appears the anti-QCD-monopole at $z=0$ for
\nextline
$-\sqrt{3}|z_0|<t<\sqrt{3}|z_0|$.

Fig.2 (b) shows an example for two instantons with
the same size, $a_1=a_2$, but a little rotated in ${\bf R}^4$ as
$(z_1,t_1)=-(z_2,t_2)$ =(1,0.05).
In this case, the QCD-monopole trajectory has a folded structure
[\sugaX,\sugamiya].
Fig.2 (c) shows an example for two instantons locating at
the same time $(z_1,t_1)=-(z_2,t_2)$=(1,0), but with a little different
size, $a_2=1.1a_1$.
There appears a QCD-monopole loop in this case [\sugaX,\sugamiya].
Thus, the QCD-monopole trajectories originating from instantons
are very unstable against a small fluctuation relating to
the location or the size of instantons [\sugaX,\sugamiya].

For a general $N$-instanton system with $N \ge 3$,
it is rather difficult to find a suitable gauge transformation
diagonalizing $A_4(x)$, and therefore it is hard to obtain
the QCD-monopole trajectory.
However, the QCD-monopole trajectory can be also
obtained by $x=y=0$ and $A_4(z,t)=0$ as Eq.{\ELEVEN} for
the multi-instantons located on the $zt$-plane: $x_k=y_k=0$.
We examine such a special case in the multi-instanton system,
where $A_4(x)$ is given by
$$
A_4(x) =
{i \over \phi (x)} \left(
(\tau _x x+\tau _y y)
\sum_k {a_k^2  \over |x-x_k|^4}
+\tau _z \sum_k {a_k^2  (z-z_k) \over |x-x_k|^4} \right).
\eqn\Aone
$$
Near the QCD-monopoles $x_s^\mu  \equiv (x_s,y_s,z_s,t_s)$
obeying $A_4(x_s^\mu )$ =0 with $x_s=y_s$=0,
$A_4(x)$ becomes the hedgehog configuration at each time $t_s$,
$$
\eqalign{
A_4(x,y,z,t_s)
&\simeq
{i \over \phi (x_s)} \biggl\{ (\tau _x x+\tau _y y)
\sum_k {a_k^2  \over |x_s-x_k|^4}
+ \tau_z f(z;t_s) \biggr\}
\cr
&\simeq
{i \over \phi (x_s)} \biggl\{ (\tau _x x+\tau _y y)
\sum_k {a_k^2  \over |x_s-x_k|^4}
+ \tau_z (z-z_s) \partial_z f(z_s;t_s) \biggr\},
}
\eqn\Atwo
$$
where the function $f(z;t_s)$ is given by
$$
f(z;t_s) \equiv \sum_k
{a_k^2  (z-z_k) \over \{(z-z_k)^2+(t_s-t_k)^2\}^2}.
\eqn\Athree
$$

We examine the correspondence between QCD-monopoles and
the function $f(z;t_s)$ at a fixed time $t_s$.
The nodes of $f(z;t_s)$ provide the QCD-monopole
trajectory in the $zt$-plane: $f(z_s;t_s)=0$.
Since the factor $\sum_k {a_k^2  \over |x_s-x_k|^4}$ is
positive definite, the magnetic charge of the QCD-monopole depends
only on the sign of the derivative $\partial_z f(z_s;t_s)$:
For $\partial_z f(z_s;t_s)>0$, $A_4(x)$ becomes the
hedgehog around $x_s^\mu $, so that the QCD-monopole appears at $x_s^\mu $.
For $\partial_z f(z_s;t_s)<0$, $A_4(x)$ becomes the
anti-hedgehog around $x_s^\mu $,
and therefore the anti-QCD-monopole with the opposite magnetic charge
appears at $x_s^\mu $.
Thus, the nodes of $f(z;t_s)$ with the positive derivative
provide the QCD-monopoles, and those with the negative derivative
provide the anti-QCD-monopoles.
We call hereafter
the node with the positive derivative as the positive node,
and that with the negative derivative as the negative node.
Since the positive and negative nodes appear alternately
in the continuous function $f(z;t_s)$,
the QCD-monopole and the anti-QCD-monopole
appear by turns spatially in the $zt$-plane.
The function $f(z;t_s)$ has a definite asymptotic form
as $z\rightarrow \pm \infty $,
$$
f(z;t_s) \sim z^{-3} \sum_k a_k^2
\eqn\Afour
$$
independent of $t_s$ and the instanton centers, $x_k^\mu $.
Hence, the number of the positive node is one more than
that of the negative node, which means that
the total magnetic charge is always unity at each time $t_s$.
The conservation law on the magnetic charge is thus guaranteed,
which was also seen in Fig.2.

In general, the QCD-monopole trajectory becomes highly complicated
and unstable in the multi-instanton system even at the classical level,
and a small fluctuation of instantons often changes the topology of
the QCD-monopole trajectory as shown in Fig.2.
In addition, the quantum fluctuation would make it more complicated
and more unstable, which leads to appearance of a long twining
trajectory as a result.
Hence, instantons may contribute to promote monopole condensation,
which is signaled by a long complicated monopole loop in the
lattice QCD simulation [\hioki,\shiba,\kitahara].

Very recently, the strong correlation between instantons and
QCD-monopoles has been also supported in the
lattice QCD simulations
\REF\thurner{
S.~Thurner, H.~Markum and W.~Sakuler,
{\it Color Confinement and Hadrons},
(World Scientific, 1995) in press.
}
[\sugaX,\sugamiya,\origuchi,\thurner], and
the monopole dominance
[\sugamiya,\hioki,\kronfeld-\giacomo]
for the topological charge has been pointed out
both in the maximally abelian gauge [\origuchi] and
in the Polyakov gauge [\sugaX,\sugamiya]
by investigating the instanton number
after the decomposition of the abelian link variable into
the singular (monopole-dominating) part and the regular
(photon-dominating) part.

\chapter{QCD-monopoles in the Instanton and Anti-instanton System}

In this chapter, we study the QCD-monopole trajectory
in the instanton and anti-instanton ($I$-${\bar I}$) system within the
dilute gas approximation.
For the instanton and the anti-instanton
located at $x_1^\mu $ and $x_2^\mu $, respectively,
one obtains
$$
\eqalign{
A^\mu (x) &=-i{\bar \eta ^{a\mu \nu }\tau ^a \over \phi _{I \bar I}(x)}
\biggl\{ {a_1^2 (x-x_1)^\nu  \over |x-x_1|^4}
-{a_2^2 (x-x_2)^\nu  \over |x-x_2|^4}\biggr\},
\cr
\phi _{I \bar I}(x)
&\equiv 1+{a_1^2 \over |x-x_1|^2}-{a_2^2 \over |x-x_2|^2}
}
\eqn\Bone
$$
within the dilute gas approximation.
Similarly to the previous chapter,
one can put the centers of the instanton and
the anti-instanton in $zt$-plane without loss of generality:
$x_1$=$y_1$=$x_2$=$y_2$=0.
The QCD-monopole trajectory only appears on the $zt$-plane,
because of the axial-symmetry around the $z$-axis of the system,
so that one has only to examine $A_4(x)$ on the $zt$-plane
by setting $x=y=0$.
In this case, $A_4(x)$ is already diagonalized on the $zt$-plane:
$$
A_4(0,0,z,t)
= {-i \tau^3  \over \phi _{I \bar I}(z,t)}
\biggl\{{a_1^2 (z-z_1) \over \{ (z-z_1)^2+(t-t_1)^2 \}^2}
-{a_2^2 (z-z_2) \over \{ (z-z_2)^2+(t-t_2)^2 \}^2} \biggr\}.
\eqn\Btwo
$$
Hence, the QCD-monopole trajectory $x^\mu =(x,y,z,t)$ in the
Polyakov-like gauge
is simply given by $x=y=0$ and $A_4(0,0,z,t)=0$ or
$$
 {a_1^2 (z-z_1) \over \{ (z-z_1)^2+(t-t_1)^2 \}^2}
-{a_2^2 (z-z_2) \over \{ (z-z_2)^2+(t-t_2)^2 \}^2}=0.
\eqn\Bthree
$$
Near the QCD-monopole, $A_4(x)$ takes a hedgehog or an anti-hedgehog
configuration at each $t$.
The centers of the instanton and the anti-instanton are
penetrated by the QCD-monopole and the anti-QCD-monopole, respectively.
In Fig.3, we show the typical cases of the $I$-${\bar I}$ system.
The QCD-monopole trajectory is found to be simple two curves
penetrating the centers of the instanton and the anti-instanton.

\chapter{QCD-monopoles in the Thermal-Instanton System}

We also study the thermal instanton system in the Polyakov-like gauge.
The multi-instanton solution at finite temperature $T$ is
given by
$$
A^\mu (x)
=i\bar \eta ^{a\mu \nu }{\tau ^a \over 2} \partial^\nu \ln \phi _{_{T}}(x)
=i\bar \eta ^{a\mu \nu }{\tau ^a \over 2} \partial^\nu \phi _{_{T}}(x)/\phi
_{_{T}}(x),
\eqn\TWELVEhalf
$$
where $\phi _{_{T}}(x)$ is a scalar function,
$$
\eqalign{
\phi _{_{T}}(x)&=1+\sum_k a_k^2 \sum_{n=-\infty }^\infty
{1 \over ({\bf x}-{\bf x}_k)^2+(t-t_k-n/T)^2} \cr
&=1+\pi T \sum_k {a_k^2 \over |{\bf x}-{\bf x}_k|}
\cdot
{{\rm sinh}(2\pi T|{\bf x}-{\bf x}_k|) \over
{\rm cosh}(2\pi T|{\bf x}-{\bf x}_k|) -\cos \{2\pi T(t-t_k)\}}.
}
\eqn\THIRTEEN
$$
In this system, $A_4(x)$ is given by
$$
\eqalign{
A_4(x)
=&-i{\pi T \tau ^a \over 2\phi _{_{T}}}\sum_k
{a_k^2 ({\bf x}-{\bf x}_k)^a \over |{\bf x}-{\bf x}_k|^3}
\bigl(
{{\rm sinh}(2\pi T|{\bf x}-{\bf x}_k|) \over
{\rm cosh}(2\pi T|{\bf x}-{\bf x}_k|) -\cos \{2\pi T(t-t_k)\}} \cr
& -2\pi T |{\bf x}-{\bf x}_k| \cdot
{1-{\rm cosh}(2\pi T|{\bf x}-{\bf x}_k|) \cos\{2\pi T(t-t_k) \} \over
[{\rm cosh}(2\pi T|{\bf x}-{\bf x}_k|) -\cos \{2\pi T(t-t_k)\}]^2}
\bigr)
}
\eqn\FOURTEEN
$$

At the high-temperature limit $T\rightarrow \infty $,
$$
A_4(x) \simeq -{i\pi T \over 2\phi _{_{T}}} \tau^a
\sum_k {a_k^2 ({\bf x}-{\bf x}_k)^a \over |{\bf x}-{\bf x}_k|^3}
\eqn\SIXTEEN
$$
becomes time-independent, so that $A_4({\bf x})$ can be diagonalized
using a time-independent gauge transformation by $\hat \Omega ({\bf x})$,
$$
A_4({\bf x})\rightarrow  \hat \Omega ({\bf x})A_4({\bf x}) \hat \Omega
^{-1}({\bf x})
=A_4^d({\bf x}),
\eqn\FIFTEEN
$$
where QCD-monopoles appear at
the points ${\bf x}_s$ satisfying $A_4({\bf x}_s)=0$,
These points ${\bf x}_s$ includes
all the centers of instantons, ${\bf x}_k$,
and become the centers of the (anti-)hedgehog configuration
on $A_4({\bf x})$.
Thus, the QCD-monopole trajectory
is reduced to simple straight lines
$$
{\bf x}={\bf x}_s \hbox{\quad} (-\infty <t<\infty ),
\eqn\SEVENTEEN
$$
where each instantons are penetrated in the temporal direction.
Such a simplification of the QCD-monopole trajectory
may corresponds to the deconfinement phase transition through
the vanishing of QCD-monopole condensation
\REF\ichie{
H.~Ichie, H.~Suganuma and H.~Toki, Phys.~Rev.~{\bf D52} (1995) 2944.
}
[\hioki,\kronfeld-\ejiri,\sugaB,\ichie].

For the thermal two-instanton system, all instanton centers
can be put on the $zt$-plane by a suitable spatial
rotation in ${\bf R}^3$
like the two-instanton system at $T=0$,
so that one can set as $x_k=y_k=0$ $(k=1,2)$.
Owing to the axial-symmetry around the $z$-axis of the system,
the QCD-monopole trajectory only appears on the $zt$-plane,
where $A_4(x)$ in Eq.{\FOURTEEN} is already diagonalized.
Hence, the QCD-monopole trajectory $x^\mu =(x,y,z,t)$
is simply given by $x=y=0$ and $A_4(z,t;x=y=0)=0$.
Here, $A_4(x)$ takes a hedgehog or an anti-hedgehog
configuration near the QCD-monopole at each $t$.

We show in Fig.4 the typical examples of the QCD-monopole trajectory
in the thermal two-instanton system.
As temperature goes high, the trajectory tends to be
straight lines in the temporal direction.
There also appears the QCD-monopole with the opposite
magnetic charge at the point satisfying $A_4(x)=0$.
The topology of the QCD-monopole trajectory is drastically changed
at $T_c \simeq 0.6 d^{-1}$, where $d$ is the distance between the
two instantons.
Since the classical Yang-Mills theory has no scale-parameter,
the typical scale of the system is mainly characterized by
the relative distance $d$.
If one adopts $d \sim 1{\rm fm}$
as a typical mean distance between instantons, such a topological
change occurs at $T_c \sim 120 {\rm MeV}$ [\sugaX,\sugamiya].

\chapter{Summary and Concluding Remarks}

We have studied the physical meanings of the abelian gauge fixing
in terms of the analogy between the nonperturbative QCD vacuum and
the superconductor.
In the abelian gauge, the QCD-monopole appears from the hedgehog
configuration corresponding to the homotopy group
$\pi _2({\rm SU}(N_c)/{\rm U}(1)^{N_c-1})=Z_\infty ^{N_c-1}$.
The ordering condition in the abelian gauge fixing
is important for the determination of the magnetic charge of
the QCD-monopoles.

We have studied the relation between instantons and monopoles
in the abelian gauge.
Simple topological consideration indicates that instantons
survive only around the world line of the QCD-monopole, which is
the topological defect.

We have found a close relation between instantons
and the QCD-monopole trajectory
in the Polyakov-like gauge, where $A_4(x)$ is to be diagonalized.
Every instantons are penetrated by the world lines of QCD-monopoles.
Anti-instantons are penetrated by the anti-QCD-monopoles
with the opposite magnetic charge.
The QCD-monopole trajectory in ${\bf R}^4$
tends to be folded and complicated
in the multi-instanton system,
although it becomes a simple straight line
in the single-instanton solution.
The QCD-monopole trajectory is very unstable against a small
fluctuation on the location and the size of instantons.
We have found that the magnetic charge conservation is
guaranteed by the argument on the node property
of a continuous function relating to $A_4(x)$.

We have studied the QCD-monopole trajectory in
the instanton and the anti-instanton system in the
Polyakov-like gauge.
The QCD-monopole trajectory becomes simple two curves penetrating
the instanton and the anti-instanton.

We have also studied the thermal instanton system in the
Polyakov-like gauge.
At the high-temperature limit, the QCD-monopole trajectory
becomes simple straight lines in the temporal direction.
The QCD-monopole trajectory drastically changes its topology
at a high temperature.

\vskip1cm

\ack{
We would like to thank Prof.~O.~Miyamura and Prof.~T.~Kanki
for useful discussions. We also thank all the members of
the theoretical group of Research Center for Nuclear Physics.
}

\refout

\FIG\FONE{
The QCD-monopole trajectory
(a) in the single-instanton system,
(b) in the single anti-instanton system.
The (anti-)instanton is denoted by a small circle.
}

\FIG\FTWO{
Examples of the QCD-monopole trajectory
in the two-instanton system with
(a) $(z_1,t_1)=-(z_2,t_2)$ =(1,0), $a_1=a_2$;
(b) $(z_1,t_1)=-(z_2,t_2)$ = (1,0.05), $a_1=a_2$;
(c) $(z_1,t_1)=-(z_2,t_2)=(1,0), a_2=1.1a_1$.
}

\FIG\FTHREE{
The QCD-monopole trajectory in the system of
the instanton and anti-instanton located
at $(0,0,z_1,t_1)$ and $(0,0,z_2,t_2)$, respectively.
The parameters are taken as
(a) $(z_1,t_1)=-(z_2,t_2)$ =(1,0), $a_1=a_2$;
(b) $(z_1,t_1)=-(z_2,t_2)$ =
(${1 \over \sqrt{2}}$,${1 \over \sqrt{2}}$), $a_1=a_2$;
(c) $(z_1,t_1)=-(z_2,t_2)=(1,0), a_2=2a_1$.
}

\FIG\FFOUR{
The QCD-monopole trajectory
in the thermal two-instanton system
with $(z_1,t_1)=-(z_2,t_2)=$ $(d/2,0)$
and $a_1=a_2$ (a) at $T^{-1}=2d$; (b) at $T^{-1}=1.5d$.
The same with $(z_1,t_1)=-(z_2,t_2)=(d/2,0.05d/2)$
and $a_1=a_2$ (c) at $T^{-1}=2d$; (d) at $T^{-1}=1.5d$.
The drastic change of the QCD-monopole trajectory is found
between the low-temperature ($T^{-1}=2d$) and
the high-temperature ($T^{-1}=1.5d$) cases.
}

\figout

\end